\documentclass{aastex}

\begin{document}

\title{X-RAY HALOS AND LARGE GRAINS IN THE DIFFUSE INTERSTELLAR MEDIUM}

\author{Adolf N. Witt}
\affil{Ritter Astrophysical Research Center, The University of Toledo,
2801 W. Bancroft, Toledo, OH 43606; awitt@dusty.astro.utoledo.edu}

\and

\author{Randall K. Smith}
\affil{High Energy Astrophysics Division, Smithsonian Astrophysical
Observatory, 60 Garden Street, Cambridge, MA 02138; rsmith@cfa.harvard.edu}

\and

\author{Eli Dwek}
\affil{Laboratory for Astronomy and Solar Physics, NASA Goddard Space
Flight Center,
Greenbelt, MD 20771; edwek@stars.gsfc.nasa.gov}

\begin{abstract}
Recent observations with dust
detectors on board the interplanetary spacecraft Ulysses and Galileo
have  recorded a substantial
flux of large interstellar grains with radii between 0.25 and 2.0
$\mu$m entering the solar system
from the local interstellar  cloud. The
most commonly used
interstellar grain size distribution is characterized by a $a^{-3.5}$
power law in grain radii
$a$, and extends to a maximum grain radius of 0.25 $\mu$m. The 
extension of the interstellar
grain size distribution to
such large radii will have a major effect on the median grain size, and on the
amount of mass needed to be tied up in dust for a given visual
optical depth. It is therefore
important to investigate whether this population of larger dust
particles prevails in the general
interstellar medium, or if it is merely a local phenomenon. The
presence of large interstellar
grains can be mainly inferred from their effect
on the intensity  and radial profile of scattering halos around X-ray
sources. In this paper we
examine the grain size distribution that gives rise to the X-ray halo
around Nova Cygni 1992. The
results of  our study confirm the need to extend the interstellar
grain size distribution in the
direction of this source to and possibly beyond  2
$\mu$m. The model  that gives the best fit to the halo data is
characterized by: (1) a grain size
distribution that follows an $a^{-3.5}$ power law up to 0.50 $\mu$m, followed
by an $a^{-4.0}$ extension from 0.50 $\mu$m to 2.0 $\mu$m; and (2)  silicate
and graphite (carbon) dust-to-gas mass ratios of 0.0044 and 0.0022,
respectively,
consistent with solar abundances constraints. Additional observations of X-ray
halos probing other spatial directions are badly needed to test the general
validity of this result.

Subject Headings: ISM: abundances -- ISM: dust, extinction -- ISM: scattering
                   X-rays: ISM
\end{abstract}

\section{INTRODUCTION}

	Dust particle detectors on space probes currently exploring the outer
solar system, e.g. Ulysses and Galileo, have provided direct evidence
of interstellar grains entering the solar system on hyperbolic
trajectories (Baguhl et al. 1996). The flow direction of these grains
closely coincides with the flow direction of the neutral gas as
determined by the Ulysses neutral gas experiment (Witte et al. 1996)
and by observations of the backscattered $He^{ 0}$ 584 \AA\ radiation (Flynn
et al. 1998). These facts suggest that the detected grains are part of
the interstellar dust population of the local interstellar cloud. An
unexpected characteristic of the in-situ detected interstellar grains
is that their inferred size distribution extends well beyond the
maximum sizes usually adopted for interstellar grains (Landgraf et
al. 2000). It must be noted that the dust detectors are able to record
the mass, speed, and direction of incoming grains with radii above 
0.05 $\mu$m, where the
grain radii are inferred by assuming spherical shapes and a bulk density
of 2.5 g cm$^{-3}$ for the particles. Most of the mass of the
observed grains is contained in particles with radii in excess of 0.35 $\mu$m
and extending to 2.0 $\mu$m, while the maximum size of of the commonly adopted
MRN size distribution (Mathis et al. 1977) for typical interstellar grains is
0.25 $\mu$m.

   Frisch et al. (1999) compared the dust mass density contained in 
this extended grain size
distribution with the gas mass density of the local interstellar gas, 
and derived a local
dust-to-gas mass ratio of $R_{\frac{d}{g}}$ = $1.1{+0.30 \atop -0.35}\times
10^{-2}$. The local dust-to-gas mass ratio derived by Frisch et al. 
represents a lower limit, since it has not been
corrected for the missing flux of smaller grains filtered out by the
heliopause,  nor has it been
corrected for the potential flux from still larger grains which are
beyond the limit of
detectability of the current detectors. Its value is about twice the 
canonical average Galactic value of
0.6 $\times 10^{-2}$ (Spitzer 1978), derived on the basis of average 
UV extinction per kpc and average gas densities, but comparable with
0.94$\times 10^{-2}$, the total
mass fraction of refractory elements present in a gas of solar system
abundances.

Independent evidence for
the existence of large interstellar grains in the size domain detected by
Ulysses and Galileo is provided from
the analysis of pre-solar grains in primitive meteorites (Anders \&
Zinner 1993;
Amari et al. 1994). Silicon carbide and graphite grains with isotopic
compositions pointing toward a supernova origin are the dominant
ones in the collection
of large presolar grains. Theoretical calculations show that the
radioactive environment of
supernova ejecta leads preferentially to the formation of relatively
large, i.e. micrometer-sized, grains (Clayton et al. 1999). However,
the total mass fraction of
interstellar dust of supernova origin is not well known. If supernova
condensates produce
most of the silicates and a significant fraction of the interstellar
carbon dust (Dwek
1998), then large interstellar grains may be prevalant in the general ISM.
Unfortunately, the present interplanetary dust detectors on board Ulysses and
Galileo cannot provide any information on the chemical
composition of the large grains to confirm these theoretical assertions.

In this Letter, we examine whether the population of large
interstellar grains is only present
in the local environment due to some local spatial or temporal
fluctuation in the grain size
distribution, or whether this population is present in the general
interstellar medium (ISM).
These large dust particles are grey at optical wavelengths, and their
existence could
therefore not be constrained by the extinction analysis of Mathis et
al. (1977). We show that the
profile and intensity level of scattered X-ray halos are potentially
the most sensitive indicator
of the relative number of large grains along interstellar
lines-of-sight (\S2). In \S3 we use the
X-ray halo of Nova Cygni 1992 and the Ulysses and Galileo constraints
on the abundance and
size distribution of large interstellar grains to study the grain
size distribution
along the Nova Cygni line of sight (\S 3). The results of our paper
and their astrophysical
implications are discussed in \S4.

\section{X-RAY HALOS AS TRACERS OF LARGE INTERSTELLAR GRAINS}

Only a few of the observable interstellar dust characteristics show
any specific sensitivity to the relative number of larger
grains. These include the wavelength of peak linear
interstellar polarization (e.g. Messinger et al. 1997), the dust
albedo at near-infrared wavelengths (Kim et al. 1994; Witt et
al. 1990; Lehtinen \& Mattila 1996; Witt et al. 1994), and the shape
and intensity of X-ray halos surrounding point sources of X-rays
(Mauche \& Gorenstein 1986; Mathis \& Lee 1991; Predehl \& Klose 1996).
Linear interstellar polarization is observed
to peak at wavelengths in the visible or near-infrared along different
lines-of-sight, and a close correlation between these peak
wavelengths and the maximum size of the aligned grains is predicted by
interstellar extinction models. Longer peak wavelengths are found
mostly in denser clouds (Messinger et al. 1997), pointing toward
larger grains there. However, for the wavelength of maximum
polarization to be a unique indicator of maximum grain size, grains of
all sizes must be aligned with equal efficiency in all
environments, and grains of all sizes must be nonspherical to the
same degree. Existing data, which relate the wavelength of maximum
polarization and the ratio of total to selective extinction, an
alignment-independent indicator of grain size, exhibit considerable
scatter (Whittet 1992; Martin et al. 1999). With the alignability of
grains of different size remaining a question, interstellar polarization
does not appear to be the most reliable indicator of grain size.

The dust albedo varies as
$a^{3}$ for particles satisfying the condition that 2$\pi a/\lambda 
\ll$ 1 (Bohren \& Huffman 1983).
For the interstellar grains detected by Ulysses and
Galileo, this wavelength region is in the near-infrared. Observations
of high-albedo scattering at near-IR wavelengths (Lehtinen \& Mattila
1996; Witt et al. 1994) have therefore been interpreted in terms of
larger than average grain sizes. Unfortunately, only relatively dense
dust structures have sufficient optical depth in the near-infrared to
produce easily measurable scattered light intensities, and, as a
result, this method is not practical as a means to study the large
end of the dust size distribution in the diffuse interstellar medium.

The shape and intensity of X-ray halos surrounding X-ray point 
sources are strongly dependent upon
the largest grains along the line of sight (Catura 1983; Bode et 
al. 1985), providing the most promising method for probing the 
abundance of large grains in the general ISM. The
efficiency for X-ray scattering varies approximately as $a^{4}$ (Catura 1983)
while the differential cross section varies as $a^{6}$ (Hayakawa 1970).
Therefore, for any typical size distribution n(a) $\sim a^{-p}$ with
3.0 $< p <$ 3.5, the largest grains will dominate the X-ray scattering halo.
Only in the weak
outer parts of the predicted halos is the effect of smaller grains
noticeable (Mathis and Lee 1991). The X-ray halo approach has the
added advantage that typical Galactic sightlines with N(H) $\sim 10^{21}$
cm$^{-2}$ are sufficiently thin to the interaction of dust with X-rays so
that single-scattering approaches are adequate, yet sufficiently long
to produce X-ray halos
intense enough to be observable with current orbiting X-ray telescopes.

In the following sections, we will briefly review the available observational
data for the sight line towards Nova Cygni 1992, and describe the
dust models used to model
the halo intensity and profile.

\section{THE X-RAY HALO OF NOVA CYGNI 1992}

\subsection{Observational Data}

The data for the X-ray intensity of the Nova Cygni 1992 halo have been
taken from Mathis et al. (1995) and are based on a 2240 s integration
with the ROSAT position sensitive proportional counter taken on 1992,
December 6, 291 days after outburst. We adopt as the hydrogen
column density a value of N(H) = 2.1 $\times 10^{21}$ cm$^{-2}$, which is based
on the mean interstellar reddening value E(B-V) = 0.36 $\pm$ 0.04
determined for
Nova Cygni 1992 by Austin et al. (1996). These values are
about 50\% higher than the corresponding values adopted by Mathis et al. (1995)
and by Smith and Dwek (1998). We consider the Austin et al. (1996) reddening
result to be more nearly definitive, given that it is based on an exhaustive
study of all optical and UV extinction indicators available for Nova
Cygni 1992.
Since the derived hydrogen column density is reddening-based, we must consider
the dust-to-gas mass ratio an adjustable parameter when we adopt a size
distribution including larger grains, because the larger grains
contribute little reddening but considerable mass.

\subsection{Adopted Dust Abundance and Size Distributions}

In the computations for this paper, we compare the scattering halos
produced by three
different grain size distributions to observations. We also examine
the effect of the spatial
distribution of the dust
along the line-of-sight on the halo.

We assumed that all the interstellar silicate, magnesium, iron is locked up
in silicate dust particles with a total atomic weight of 172 amu (Si
+ Mg + Fe + 4$\times$O), and
a mass density of 3.3 g cm$^{-3}$. For a gas of solar abundances, the
resulting silicate-to-hydrogen
mass ratio is 0.0062. We assumed that 270 carbon per 10$^6$ H atoms
are incorporated into the
dust. With a graphite mass density of 2.2 g cm$^{-3}$, the resulting
graphite-to-hydrogen mass
ratio is 0.0032. With a H/He number ratio of 10, the total
dust-to-gas mass ratio is 0.0066, close
to the canonical Galactic value of 0.006.

Three different grain size distributions, all normalized to the same
dust-to-gas mass ratio were
used in the calculations. The first consisted of the standard MRN
size distribution (Mathis et
al. 1977) that follows a $a^{-3.5}$ power law in grain radii between
$a_{min}$ = 0.005 $\mu$m,
and $a_{max}$ = 0.25 $\mu$m. The second grain size distribution was
characterized by a
$a^{-3.5}$ power law between 0.005 to 0.5 $\mu$m, followed by an
$a^{-4}$ extension
from 0.5 $\mu$m to 2.0 $\mu$m, as observed by Landgraf et al. (1999)
for interstellar
grains currently flowing through the solar system. We will refer to
this modified MRN distribution as the XMRN
distribution. As a third case, we adopted another modified MRN
distribution, this one with
the original values for the upper and lower size limits of $a_{min}$ =
0.005 $\mu$m and $a_{max}$ = 0.25 $\mu$m but with a less steep power law,
namely $a^{-2.5}$. This places a larger fraction of the mass into the larger
grains without actually changing the range of the size distribution.

Two line-of-sight dust distributions were considered: a uniform distribution
as adopted previously by Mathis et al. (1995) and Smith and Dwek (1998),
and a distribution where the entire dust is contained in the last 1/3 of
the (3.2 $\pm$ 0.5) kpc (Paresce 1994) optical path, closest to the sun.
We consider the latter distribution to be more realistic, given that the
galactic latitude of Nova Cygni 1992 of b = +7.8$^\circ$ places the source at
a z-distance of about 450 pc, well above the average Galactic interstellar dust
layer.

\subsection{Halo Predictions}

For the computation of the X-ray differential cross sections we employed
the Mie theory approach as suggested by Smith \& Dwek (1998, 2000).
This avoided
the demonstrated shortcomings of the traditionally used
Rayleigh-Gans approximation (Bohren \& Huffman 1983), which would have
substantially overestimated the scattering cross sections for the energy
of the observed X-rays of 0.4 keV (Mathis et al. 1995) and the larger grains
in our adopted size distributions. The optical constants for the
graphite and silicate
grains were taken  from Martin \& Rouleau (1991) and Rouleau \& Martin (1991).
The Mie calculations were carried out with a code developed by Wiscombe
(1979, 1980). The code was originally developed for the purpose of
modeling the optical
properties of aerosol particulates in the Earth's atmosphere, and is
therefore particularly
suited for cases in which the grain size parameter, 2$\pi a/\lambda$,
is significantly larger (maximum values up to 20,000)
than unity. For calculating the halo intensity, we adopted a point
source flux of 6.62
$\times 10^{4}$ counts at energy of 0.4 keV ($\lambda$ = 0.003
$\mu$m) and a
hydrogen column density  N(H) = 2.1$\times 10^{21}$ cm$^{-2}$.

Fig. 1 compares the calculated halo intensity and profile for the
various cases with the
observations.
Four different model curves are shown in the figure. The standard MRN
distribution
(solid curve) for a uniform dust distribution along the entire line
of sight leads to a predicted halo which is too bright by about a factor
two, except in the innermost region.
The lack of agreement with the data in the outer parts of the halo suggests
that too much of the dust mass in this model is concentrated in the
smaller grains
which dominate the outer parts of the halo. Bode et al. (1985) noted a
qualitatively similar disagreement between a predicted halo profile
for an MRN-type distribution with a maximum grain radius of 0.2 $\mu$m and
the observed halo of Cyg X-1 (Image H821). By contrast, the XMRN distribution
for the uniform dust distribution
(dotted curve) provides a much better fit. Most of the mass is now contained in
the larger grains providing the large-size extension, thus reducing the number
of smaller grains. Consequently, the innermost part of the halo
dominated by the
largest grains is enhanced while the outer halo is correspondingly weakened,
more nearly in agreement with the observations. The third size distribution
tested (dashed curve), one with a MRN upper size limit of 0.25 $\mu$m but a
less steep power law, exhibits a profile similar to that of the original
MRN distribution (solid curve) but overall larger intensities, resulting in
an unacceptable fit. As the previous two, this halo also assumes a uniform
dust distribution along the line of sight.

The fourth curve (dot-dash)  shows the predicted halo for the XMRN distribution
for a line of sight in which all grains are concentrated in the last third of
the light path closest to the observer. This curve demonstrates the
fact that in
addition to the largest grains, the grains closest to the source also add much
power to the core of the scattering halo. By moving these grains closer to
the observer the central scattering peak loses power accordingly. This fit is
qualitatively about as good as the fit of the XMRN curve with a uniform dust
distribution. It could be improved by extending the size distribution to still
larger sizes while keeping the dust-to-gas ratio constant at the
original value of 0.0066. As
demonstrated  with the examples of the three uniform line-of-sight
distributions (solid,
dotted, and dashed lines), the addition of still larger grains would
steepen the
dot-dashed profile in the central core and lower the intensity in the outer
halo, as required by the observations.

\begin{figure}
\begin{center}
\figurenum{1}
\plotone{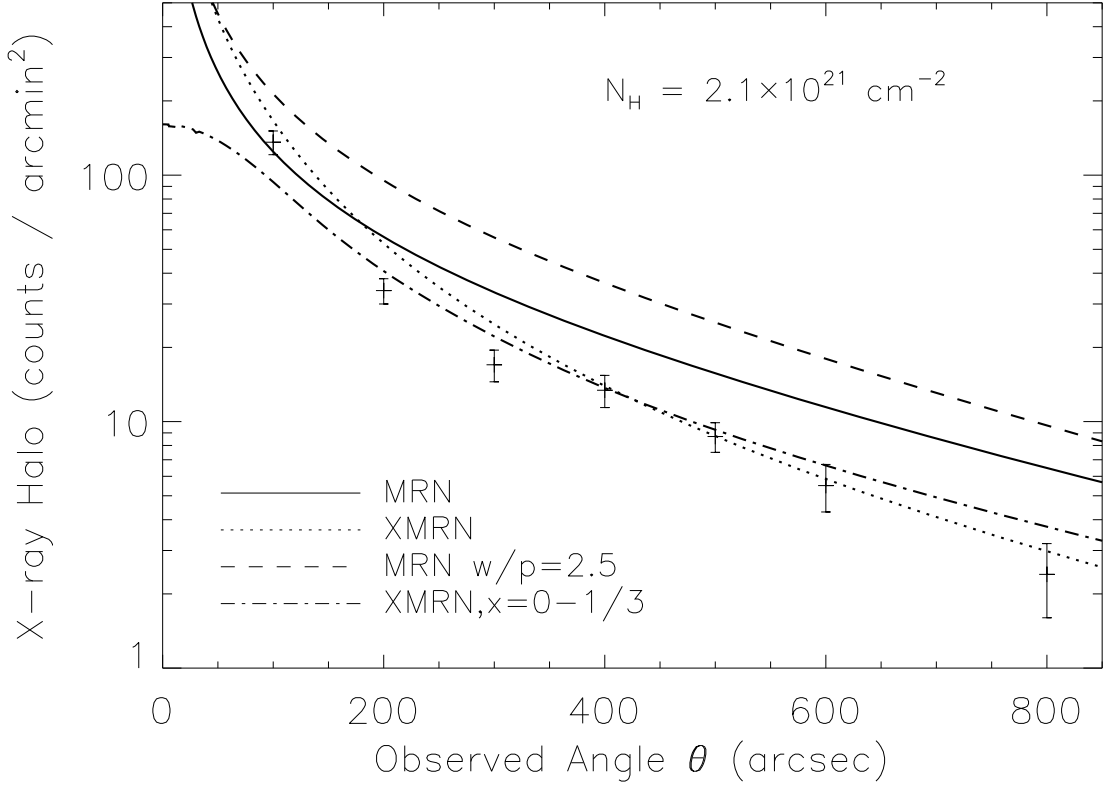}
\caption{Nova Cygni 1992 data with X-ray halo profiles predicted from Mie
theory for different size distributions and different line-of-sight dust
distributions, as described in the text. All size distributions are
normalized to yield the
canonical dust-to-gas mass ratio of 0.0066.
The MRN curve (solid) corresponds to the standard MRN distribution
of bare silicate and graphite grains. XMRN (dotted) represents a MRN
distribution extended to 2.0
$\mu$m. The dashed profile represents a size distribution
with conventional MRN size limts but a less steep power law. All three profiles
assume a uniform dust distribution along the line-of-sight. The
dot-dashed curve
represents a different line-of-sight distribution for the XMRN size
distribution , with all dust being contained in the last third
of the light path closest to the observer.}
\end{center}
\end{figure}

\section{DISCUSSION}

The results of our calculations show that extending the MRN grain
size distribution to larger
grain sizes, as characterized by the XMRN model, results in a better
fit to the X-ray halo of
Nova Cygni 1992. This suggests that particles with sizes larger than
0.25 $\mu$m constitute
a significant mass fraction of the interstellar dust population along
one line of sight through the
general ISM. Our results are therefore the first extension of the
conclusions of Frisch et al.
(1999) beyond the local Galactic neighborhood of the solar
system. However, in contrast to the
locally-based result of Frisch et al. (1999), we do not find any
evidence that the dust-to-gas
mass ratio along the Nova Cygni 1992 line-of-sight must be larger
than the canonical interstellar
value.

We need to examine to what extent our conclusions depend
on the specific dust model used in the calculations. It is possible that other
functional forms of the grain size distribution, different dust
compositions, or
different dust morphology may provide an improved fit to the Nova
Cygni 1992 halo.

There are currently a number of interstellar dust models that satisfy the usual
observational constraints with different grain size distribution (Mathis et al.
1977, Kim \& Martin 1995, Weingartner \& Draine 2000, Witt 1999). All size
distributions have
a  large range of grain sizes, they are either continuous or in a
multi-mode distribution, and most of the mass is contained in the largest
grains. No grain
size distribution postulated for the diffuse interstellar medium
incorporates large
grains in the numbers specified in Sect. 3.2, and all would require similar
modifications as we carried out with the MRN distribution in order to produce a
satisfactory fit to the X-ray halo of Nova Cygni 1992. We conclude
from this that
our result is not a peculiar consequence of the assumed size distribution.

All models utilize similar amounts of the refractory elements (Snow \&
Witt 1995,1996).The optical properties of the MRN model are
determined by the assumed
composition in terms of silicate and graphite grains with densities
specified in Sect. 3.2 and optical constants as given by Draine and
Lee (1984). As shown by Mathis et al. (1995), adopting different
compositions, especially for the carbonaceous component of the grains,
as well as composite and porous structures instead of homogeneous,
compact grains produces some variation in the overall halo intensity,
but it does not significantly affect the profile shape. This was confirmed by
the Mie-calculations for porous particles by Smith \& Dwek (1998) as well. We
conclude from this that our result concerning the need for a grain size
distribution extended to larger sizes is independent of the assumed
composition, given the normal constraints, mainly from depletion studies and
the study of grain band emissions and absorptions.

The presence of larger grains in our XMRN distribution beyond the MRN limit of
0.25 $\mu$m  does not seem to impose an additional mass requirement beyond that
set by the canonical Galactic dust-to-gas ratio, and thus our result does not
further compound the problem of the apparent mismatch between the amounts of
refractory elements required by grain models and the amounts of these elements
available in the diffuse ISM, when recent depletion measurements and B-star
reference abundances are combined (Snow \& Witt 1996). On the other hand, it
also does not solve this problem; it simply shifts more of the needed mass into
larger grains. The full discussion of this problem goes beyond the
scope of this
paper, and we refer to Frisch et al. (1999) for further details.

A serious problem does arise with respect to the XMRN distribution when one
considers its impact on the wavelength dependence of extinction. The original
MRN distribution (Mathis et al. 1977) was constructed with the constraint that
the average Galactic extinction law with $R_{V}$ = 3.2 as well as the canonical
ratio of the extinction optical depth per hydrogen atom would be matched.
By shifting most of the dust mass into grains with larger sizes, we find
that the XMRN distribution leads to an extinction curve with $R_{V}$ = 6.1
and an A${_{_V}}$/N(H) ratio of only half
the canonical value of 5.3$\times 10^{-22}$ cm$^{2}$. By contrast, 
the extinction for XMRN longward of
1~$\mu$m is  greatly enhanced  over that expected for the standard
MRN distribution.
The extinction law toward Nova Cygni 1992 is not known specifically,
and a value of
$R_{V}$ = 6.1 would suggest a peculiar grain size distribution that
differs from that
in the average ISM towards that line of sight. A more likely solution
to the present
dilemma seems to be in the high probability that neither MRN nor XMRN
with their
assumptions of chemically homogeneous, spherical solid particles are
the correct
descriptions of interstellar grains and their size  distribution. The apparent
incompatibility between the constraints arising from the X-ray halo
of Nova Cygni
1992 and the Galactic extinction law simply  illustrates again the
incompleteness of
current grain models (Witt 1999), which will not be resolved here.

We have demonstrated with the example of Nova Cygni 1992 that the line-of-sight
density distribution of the scattering dust is also a critical factor in
determining the shape of the scattering halo. We suggest, based on Nova Cygni
1992's height above the Galactic plane that a distribution with the dust
concentrated nearer to the observer is a more likely one than a uniform
distribution, but with the available data and only a single line-of-sight, this
argument lacks power. It is, therefore, essential to acquire
additional data of X-ray scattering halo profiles of other point
sources. Only when a representative sample of X-ray halo data yields
results consistent with the present finding, can the conclusions
regarding the general existence of larger grains in the diffuse
interstellar medium be considered substantiated.

Acknowledgement: We acknowledge the constructive comments received from
an anonymous referee. ANW derived material support during the period of this
work through NASA Grant NAG5-9202, which is gratefully acknowledged.
RKS acknowledges support from NASA contract NAS8-39073 and NASA GRANT
NAG5-3559.
ED acknowledges the NASA NRA 99-OSS-01 Astrophysics Theory Program for partial
support of this work.

\end{document}